\documentclass[aps,prl,superscriptaddress,amsfonts,amsmath,amssymb,showpacs,floatfix,reprint,longbibliography,footinbib]{revtex4-2}

\usepackage{url}
\usepackage{bm}
\usepackage{graphicx}
\usepackage{amsmath}
\usepackage{amstext}
\usepackage{amssymb}
\usepackage{amsfonts}
\usepackage{amsbsy}
\usepackage{verbatim}
\usepackage{color}
\usepackage[colorlinks=true, urlcolor=blue, linkcolor=blue, citecolor=blue, pdftex]{hyperref}
\usepackage[capitalise]{cleveref}
\usepackage{multirow}
\usepackage{floatrow}
\usepackage{float}
\usepackage{tikz}
\usepackage{gensymb}
\usepackage{textcomp}
\usepackage{enumitem}
\usepackage[version=3]{mhchem}
\usepackage{afterpage}
\usepackage{pbox}
\usepackage{makecell}
\usepackage{dsfont}
\usepackage{siunitx}
\usepackage{fancyhdr}
\usepackage{stackengine,wasysym,scalerel}
\usepackage{latexsym}
\usepackage{cancel}
\usepackage{array, multirow, bigdelim, makecell, booktabs}
\usepackage[misc]{ifsym}
\usepackage{times}

\newcommand{\sktrue}{\ce{Na6Cu7BiO4(PO4)4[Cl{,}(OH)]3}}
\newcommand{\skag}{\ce{Na6Cu7BiO4(PO4)4Cl3}}

\newcommand{\Cuthree}{Cu(3) }

\usepackage{array, multirow, bigdelim, makecell, booktabs} 

\begin{document}
\title{Quantum paramagnetism in the decorated square-kagome antiferromagnet {\skag}}

\author{Nils Niggemann}
\thanks{These authors contributed equally.}
\affiliation{Dahlem Center for Complex Quantum Systems and Fachbereich Physik, Freie Universit\"at Berlin, 14195 Berlin, Germany}
\affiliation{Helmholtz-Zentrum Berlin für Materialien und Energie, Hahn-Meitner-Platz 1, 14109 Berlin, Germany}
\affiliation{Department of Physics and Quantum Centre of Excellence for Diamond and
Emergent Materials (QuCenDiEM), Indian Institute of Technology Madras, Chennai 600036, India}

\author{Nikita Astrakhantsev}
\thanks{These authors contributed equally.}
\affiliation{Department of Physics, University of Z\"urich, Winterthurerstrasse 190, CH-8057 Z\"urich, Switzerland}

\author{Arnaud Ralko}
\thanks{These authors contributed equally.}
\affiliation{Institut Néel, UPR2940, Université Grenoble Alpes, CNRS, Grenoble, FR-38042 France}
\affiliation{Department of Physics and Quantum Centre of Excellence for Diamond and
Emergent Materials (QuCenDiEM), Indian Institute of Technology Madras, Chennai 600036, India}

\author{Francesco Ferrari}
\thanks{These authors contributed equally.}
\affiliation{Institut f\"ur Theoretische Physik, Goethe Universit\"at Frankfurt, Max-von-Laue-Stra{\ss}e 1, 60438 Frankfurt am Main, Germany}
\affiliation{Department of Physics and Quantum Centre of Excellence for Diamond and
Emergent Materials (QuCenDiEM), Indian Institute of Technology Madras, Chennai 600036, India}

\author{Atanu Maity}
\affiliation{Department of Physics and Quantum Centre of Excellence for Diamond and
Emergent Materials (QuCenDiEM), Indian Institute of Technology Madras, Chennai 600036, India}

\author{Tobias M\"uller} 
\affiliation{Institut f\"{u}r Theoretische Physik und Astrophysik, 
  Julius-Maximilians-Universit\"at W\"{u}rzburg, Am Hubland, D-97074 W\"{u}rzburg, Germany}

\author{Johannes Richter}
\affiliation{Institut für Physik, Otto-von-Guericke-Universität Magdeburg, P.O. Box 4120, 39016 Magdeburg, Germany}
\affiliation{Max-Planck-Institut für Physik Komplexer Systeme, Nöthnitzer Straße 38, D-01187 Dresden, Germany}

\author{Ronny Thomale} 
\affiliation{Institut f\"{u}r Theoretische Physik und Astrophysik, 
  Julius-Maximilians-Universit\"at W\"{u}rzburg, Am Hubland, D-97074 W\"{u}rzburg, Germany}
\affiliation{Department of Physics and Quantum Centre of Excellence for Diamond and
Emergent Materials (QuCenDiEM), Indian Institute of Technology Madras, Chennai 600036, India}

\author{Titus Neupert}
\affiliation{Department of Physics, University of Z\"urich, Winterthurerstrasse 190, CH-8057 Z\"urich, Switzerland}

\author{Johannes Reuther}
\affiliation{Dahlem Center for Complex Quantum Systems and Fachbereich Physik, Freie Universit\"at Berlin, 14195 Berlin, Germany}
\affiliation{Helmholtz-Zentrum Berlin für Materialien und Energie, Hahn-Meitner-Platz 1, 14109 Berlin, Germany}
\affiliation{Department of Physics and Quantum Centre of Excellence for Diamond and
Emergent Materials (QuCenDiEM), Indian Institute of Technology Madras, Chennai 600036, India}

\author{Yasir Iqbal}
\affiliation{Department of Physics and Quantum Centre of Excellence for Diamond and
Emergent Materials (QuCenDiEM), Indian Institute of Technology Madras, Chennai 600036, India}

\author{Harald O. Jeschke}
\affiliation{Research Institute for Interdisciplinary Science, Okayama University, Okayama 700-8530, Japan}
\affiliation{Department of Physics and Quantum Centre of Excellence for Diamond and
Emergent Materials (QuCenDiEM), Indian Institute of Technology Madras, Chennai 600036, India}

\begin{abstract}
The square-kagome lattice Heisenberg antiferromagnet is a highly frustrated Hamiltonian whose material realizations have been scarce. We theoretically investigate the recently synthesized {\skag} where a Cu$^{2+}$ spin-$1/2$ square-kagome lattice (with six site unit cell) is decorated by a seventh magnetic site alternatingly above and below the layers. The material does not show any sign of long-range magnetic order down to 50 mK despite a Curie-Weiss temperature of $-212$ K indicating a quantum paramagnetic phase. Our DFT energy mapping elicits a purely antiferromagnetic Hamiltonian that features longer range exchange interactions beyond the pure square-kagome model and, importantly, we find the seventh site to be strongly coupled to the plane. We combine two variational Monte Carlo approaches, pseudo-fermion/Majorana functional renormalization group and Schwinger-Boson mean field calculations to show that the complex Hamiltonian of {\skag} still features a nonmagnetic ground state. We explain how the seventh Cu$^{2+}$ site actually aids the stabilization of the disordered state. We predict static and dynamic spin structure factors to guide future neutron scattering experiments.
\end{abstract}

\date{\today}

\maketitle

{\it Introduction}. Magnetic ions forming the kagome lattice, a corner sharing network of triangles, have been the focus of several decades of highly frustrated magnetism research~\cite{Norman2016}. Kagome lattice antiferromagnets provide some of the most promising examples of highly correlated nonmagnetic ground states~\cite{Iqbal2013,He2017}, and are therefore subject of intense experimental efforts while inspiring a wealth of theoretical developments~\cite{Harrison2004,Savary2017}. Interestingly, the square-kagome lattice as a differently connected lattice of corner sharing triangles~\cite{Siddharthan2001}, can also support a quantum paramagnetic ground state~\cite{Richter-2009,Nakano-2013,Hasegawa-2018,Morita-2018,Tomczak-2003,gembe2023noncoplanar}. The precise nature of the ground state is under debate, with proposals spanning a pinwheel valence bond crystal (VBC)~\cite{Astrakhantsev2021,Rousochatzakis-2013}, length six loop VBC~\cite{Schmoll-2022,Ralko-2015}, and (lattice) nematic quantum spin liquid~\cite{Lugan-2019}. The field of frustrated quantum magnetism is currently poised with the arrival of new materials based on the square-kagome lattice geometry promising to host exotic nonmagnetic phases at low temperatures~\cite{Yakubovich2021,Liu2022,Markina2022,Murtazoev-2023}. In the most prominently studied example KCu$_6$AlBiO$_4$(SO$_4$)$_5$Cl, the Cu$^{2+}$ $S=1/2$ moments do not show any sign of long-range dipolar magnetic order down to 50 mK despite Curie-Weiss temperatures of $-237$\,K, with indications of gapless quantum spin liquid behavior~\cite{Fujihala-2020}. Recently, {\sktrue}, a novel sodium bismuth oxo-cuprate phosphate chloride containing square-kagome layers of \ce{Cu^{2+}} ions was synthesized~\cite{Yakubovich2021}. It contains, besides the six magnetic sites making up the square-kagome lattice, a seventh decorating site which is placed either above or below the square in checkerboard fashion. A study of specific heat indicates that the compound does not order magnetically down to 50\,mK~\cite{Liu2022} despite a large negative Curie-Weiss temperature of $-212$ K. The scenario in both these compounds is then strikingly similar to the kagome lattice based candidate quantum spin liquid material Herbertsmithite~\cite{Norman2016}. 

\begin{figure*}[htb]
\includegraphics[width=\textwidth]{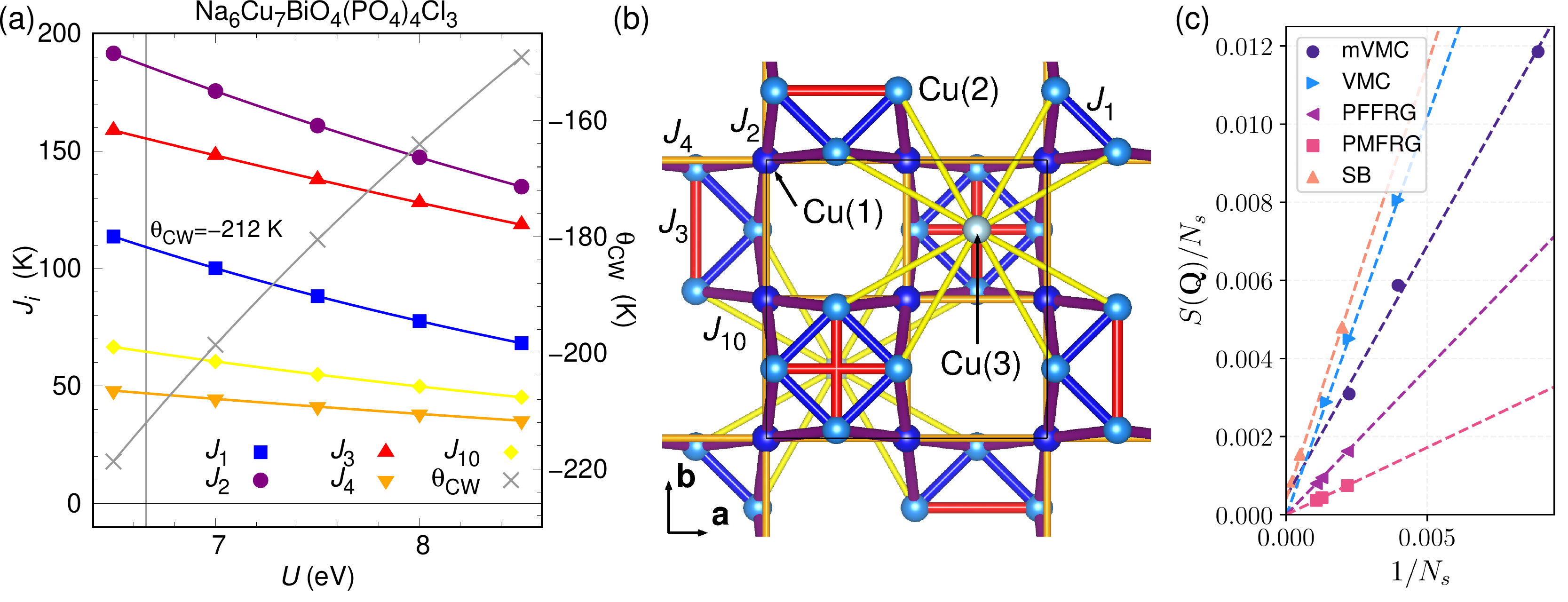}
\caption{ 
(a) Heisenberg Hamiltonian parameters of {\skag} determined by DFT energy mapping as function of onsite interaction strength $U$ (negligible couplings are not shown). The vertical line indicates the $U$ value where the exchange couplings match the experimental~\cite{Yakubovich2021} Curie-Weiss temperature. The resulting nonnegligible exchange couplings are $J_1= 109.1(8)$\,K, $J_2= 186.2(7)$\,K, $J_3= 155.3(1.4)$\,K, $J_4= 46.9(4)$\,K, $J_{10}= 64.6(2)$\,K.
(b) Relevant exchange paths of {\skag}.
(c) Finite-size scaling of the maxima of the equal-time structure factor $S(\mathbf{Q})/N_{s}$ from many-variable VMC (mVMC), VMC, pseudo-fermion FRG (PFFRG), pseudo-Majorana FRG (PMFRG) (at $T=0.2J_2$), and the Schwinger-Boson mean-field theory method. Note that, compared to the other methods, PFFRG and PMFRG use a different definition of the system size $N_s$, which counts the number of correlated sites around a reference site, likely explaining the quantitative differences of our results. Furthermore, PMFRG results are obtained at a finite temperature $T=0.2J_2$.}

\label{fig:exchange}
\end{figure*}

In this work, we will establish the Hamiltonian of {\skag} by density functional theory based energy mapping. As it is highly nontrivial to work out the ground state and excited state properties of this complex lattice with three symmetry inequivalent magnetic sites, we apply two types of variational Monte Carlo (VMC), two flavors of functional renormalization group (FRG) calculations and the Schwinger boson (SB) formalism. We establish that the Hamiltonian of {\skag} indeed realizes a nonmagnetic ground state, and provide evidence that the seventh magnetic site decorating the square-kagome lattice plays an important role in enhancing the degree of frustration, thus aiding the formation of a magnetically disordered phase in this material. This phase is shown to be a gapped VBC breaking translation symmetry, with a dimer pattern that is periodic in a $2\times2$ enlarged unit cell. We present its spectroscopic signatures to compare with future neutron scattering experiments.

{\it Heisenberg Hamiltonian}. We determined the magnetic interactions of {\skag} using all electron density functional theory calculations. We use the crystal structure determined in Ref.~\onlinecite{Yakubovich2021} but simplify it slightly by choosing the majority Na(2) position and by removing O(5) from the Cl(3) position. All 14 \ce{Cu^{2+}} ions in the primitive unit cell of the tetragonal structure are in square planar coordination with oxygen. The network they form is shown in Fig.~\ref{fig:exchange}\,(b), with the three symmetry inequivalent Cu(1), Cu(2) and \Cuthree shown in different colors. Cu(1) and Cu(2) form a square-kagome lattice, and \Cuthree is decorating this lattice above and below. Spin-polarized calculations show that the Cu$^{2+}$ ions have $S=\frac{1}{2}$ moments, and at $U=6.5$\,eV the system is insulating with a gap of $E_{\rm g}=1.7$\,eV. We use the energy mapping technique that has yielded very good results in other copper based magnets~\cite{Jeschke2015,Hering2022} to extract the Heisenberg Hamiltonian parameters. Figure~\ref{fig:exchange}\,(a) shows the result of these calculations. Exchange couplings evolve smoothly with the on-site Coulomb repulsion $U$, and the Hamiltonian reproduces the experimental Curie-Weiss temperature at $U=6.66$\,eV (vertical line). Some couplings that are less than {3\%} of the largest coupling $J_2$ are not shown in the plot. Among the couplings we resolve, there is only one, negligibly small, interlayer coupling (see Ref.~\cite{SM}); as there is a full Na and Cl layer separating the magnetic layers, we expect the deviations from magnetic two-dimensionality to be small and beyond the scope of the present study. The two couplings making up the square-kagome lattice, $J_1=0.59J_2$ and $J_2$, are the third largest and largest coupling, respectively. The diagonal in the squares, $J_3=0.83J_2$, is the second largest coupling. One second nearest neighbor of the square-kagome lattice, $J_4=0.25J_2$, is also substantial. Furthermore, the $J_{10}=0.35J_2$ interaction, which couples the square-kagome site Cu(2) to the magnetic decorating sites \Cuthree is found to be important. Strong buckling of the square-kagome lattice means that this coupling is the closest connection between \Cuthree and Cu(2) sites. It has a reasonable superexchange path through a phosphate group. On the other hand, the closer connection of a \Cuthree site to a Cu(1) site turns out to be negligible ($J_5=0.03J_2$). 

\begin{figure*}[htb]
\includegraphics[width=\textwidth]{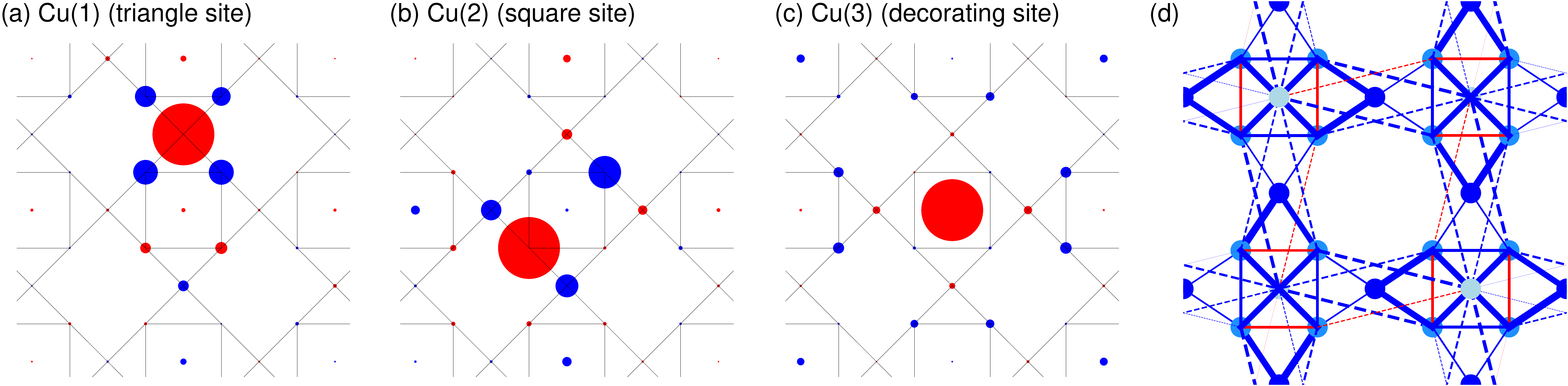}
\caption{(a)-(c) The pattern of real space (equal-time) spin-spin correlations $\langle \mathbf{\hat S}_{i}\cdot \mathbf{\hat S}_{j}\rangle$ from mVMC measured with respect to the three symmetry inequivalent sites. The radius of the circle is proportional to the magnitude of the correlator and blue (red) denote antiferromagnetic (ferromagnetic) correlations. The largest red circle corresponds to $i = j$. (d) The pattern of $\langle \mathbf{\hat S}_{i}\cdot \mathbf{\hat S}_{j}\rangle$ within a $2\times2$ unit cell showing the pattern of strong/weak bonds in the VBC ground state. The thickness is proportional to $|\langle \mathbf{\hat S}_{i}\cdot \mathbf{\hat S}_{j}\rangle|$ and blue (red) denote antiferromagnetic (ferromagnetic) bonds, while the dashed lines denote the $J_{10}$ bonds. Note that the idealized 2D unit cell shown here is rotated by $45^\circ$ with respect to Fig.~\ref{fig:exchange}(b).
}\label{fig:mvmc}
\end{figure*}

The space group $P4/nmm$ of {\skag} dictates that the $J_{1}$ (light blue) square and the $J_{2}$ (purple) triangle couplings are symmetry inequivalent as for the ideal lattice. The isotropic $J_{1}=J_{2}$ Heisenberg antiferromagnet is host to a VBC ground state with a finite spin gap $\Delta\sim0.04 J_{1}$~\cite{Schmoll-2022}, whose precise nature is still under debate~\cite{Schmoll-2022,Astrakhantsev2021,Ralko-2015,Rousochatzakis-2013}. For $J_{2}/J_{1}\geqslant2$, classically the system is host to a long-range ferrimagnetically ordered ground state (up-up-down)~\cite{Morita-2018}, however, for $S=1/2$, exact diagonalization studies on $N=24,30,36$ site clusters~\cite{Morita-2018,Richter-2023,Richter-2022} find that the system enters the ferrimagnetic ground state for $J_{2}/J_{1}\gtrsim 1.65$. The DFT estimated material couplings with $J_{2}/J_{1}\sim 1.7$ thus precariously places the system in the vicinity of the nonmagnetic-magnetic phase boundary. It is then likely that the significant $J_{3}$ diagonal couplings ($J_{3}/J_{1}\sim 1.45$) within the squares generate the necessary frustration to tip the balance in favor of a nonmagnetic ground state. If so, the precise nature of the nonmagnetic state, in the presence of further neighbor coupling $J_{4}$ and the coupling $J_{10}$ to the decorating \Cuthree site, needs to be carefully investigated by probing the delicate energetic competition between various quantum spin liquid and VBC ans\"atze.

{\it Results}. We begin our analysis by addressing the issue of the existence of long-range magnetic order in the ground state of the DFT Hamiltonian. Employing state-of-the-art numerical approaches of mVMC~\cite{misawa2019mvmc,doi:10.1143/JPSJ.77.114701}, fermionic VMC~\cite{sorella_green_1998,becca_quantum_2017}, PFFRG~\cite{Reuther2010,Iqbal-2016_3dpfrg,muller2023pseudofermion}, PMFRG~\cite{Niggemann2021,Niggemann2022}, and SB analysis~\cite{Auerbach1998,Halimeh2016,Schaffer2017,Lugan-2019,Lugan-2022}, we compute the static (equal-time) spin structure factor
\begin{equation}
    \label{eq:S_q}
    S(\mathbf q) = \frac{1}{N_{s}} \sum\limits_{0 \leqslant i,\,j < N_{s}} \langle{\mathbf{\hat S}}_i \cdot {\mathbf{\hat S}}_j \rangle e^{i\mathbf q\cdot (\mathbf r_i - \mathbf r_j)},
\end{equation}
where $N_{s}$ is the number of sites in the lattice, $\mathbf q$ is a momentum inside the extended Brillouin zone, and $\mathbf r_i$ denotes the site positions (accounting for sublattice displacements), following the convention outlined in Supp. Mat.~\cite{SM}. Long-range dipolar magnetic order sets in when the maximum of $S(\mathbf{q})$ at $\mathbf{q}=\mathbf{Q}$ scales as $S(\mathbf{Q})\propto N_{s}$ for large $N_{s}$ \cite{SandvikScaling1997}. The size scaling of $S(\mathbf{Q})/N_{s}$ [see Fig.~\ref{fig:exchange}\,(c)], yields the magnetization $m^{2}\propto {\rm lim}_{{N}_{s}\to\infty}S(\mathbf{Q})/N_{s}$, which we consistently find to be zero (within error bars) from different approaches. This provides evidence for a nonmagnetic ground state which is corroborated by the rapid decay of the real space spin-spin correlations seen in Fig.~\ref{fig:mvmc}\,(a)--(c) [see Ref.~\onlinecite{SM} for results from fermionic VMC and Schwinger boson approaches]. 

\begin{table}[b]
\begin{center}
\renewcommand{\theadfont}{\normalsize\bfseries}
\begin{tabular}{ccccl}
Unit cell & $4\times 4$ & $6\times 6$ & $8\times 8$\\
\cmidrule{1-4}
$1\times 1$ & -0.4256(1) & -0.4205(1) & -0.4172(1) & \hspace{0em}\rdelim\}{2}{*}[\parbox{2cm-\tabcolsep-\widthof{$\Bigg]$}}{With \Cuthree}] \\
$2\times 2$ & -0.4426(1) & -0.4304(1) & -0.4277(1) \\ 
\cmidrule{1-4}
\cmidrule{1-4}
$1\times 1$ & -0.4729(1) & -0.4783(1) & -0.4698(2) &\hspace{0em}\rdelim\}{2}{*}[\parbox{2cm-\tabcolsep-\widthof{$\Bigg]$}}{Without \Cuthree}] \\
$2\times 2$ & -0.4857(1) & -0.4816(1) & -0.4811(2) \\
\end{tabular}
\caption{mVMC energies $E / J_2$ on the $4\times 4$, $6\times 6$ and $8\times 8$ lattices with a $1\times 1$ and $2\times 2$ unit cell after symmetrization.}
    \label{tab:mVMCenergies}
\end{center}
\end{table}

To further elucidate the nature of the nonmagnetic ground state, we perform mVMC simulations with {\it ans\"atze} of different unit cell sizes. Table~\ref{tab:mVMCenergies} shows variational ground state energies for the {\skag} Hamiltonian on three different clusters after symmetrization. The energies without the $J_{10}$ coupling to \Cuthree sites are also given. Independent of the system size, the energies for $2\times 2$ enlarged, i.e., 24-site unit cells are slightly (about 2\%) lower compared to translation invariant states, i.e., either quantum spin liquids or lattice nematic. Other independent approaches reach similar conclusions, lending support for a translation symmetry broken ground state. These include a fermionic VMC analysis guided by different ansätze, a self-consistent fermionic mean-field analysis of different U(1) and  $\mathds{Z}_{2}$ spin liquids ~\cite{skl_psg}, as well as a Schwinger-Boson mean-field study. The pattern of real space (equal-time) spin-spin correlations is shown in Fig.~\ref{fig:mvmc}(d), which points to its VBC nature. Here, one observes a checkerboard pattern, whereby the $J_{1}$ bonds featuring ferromagnetic (antiferromagnetic) correlations are always complemented by $J_{2}$ bonds hosting strong (weak) antiferromagnetic correlations thus forming a staggered horizontal/vertical pattern. The frustrating diagonal bonds inside the squares ($J_{3}$) show the strongest (antiferromagnetic) correlations, while the Cu(2) and \Cuthree sites are also found to be strongly correlated via $J_{10}$ bonds, the latter highlighting the decorated nature of the lattice geometry. The VBC pattern possesses only $C_{2}$ symmetry.

\begin{figure}[t]
\includegraphics[width=\columnwidth]{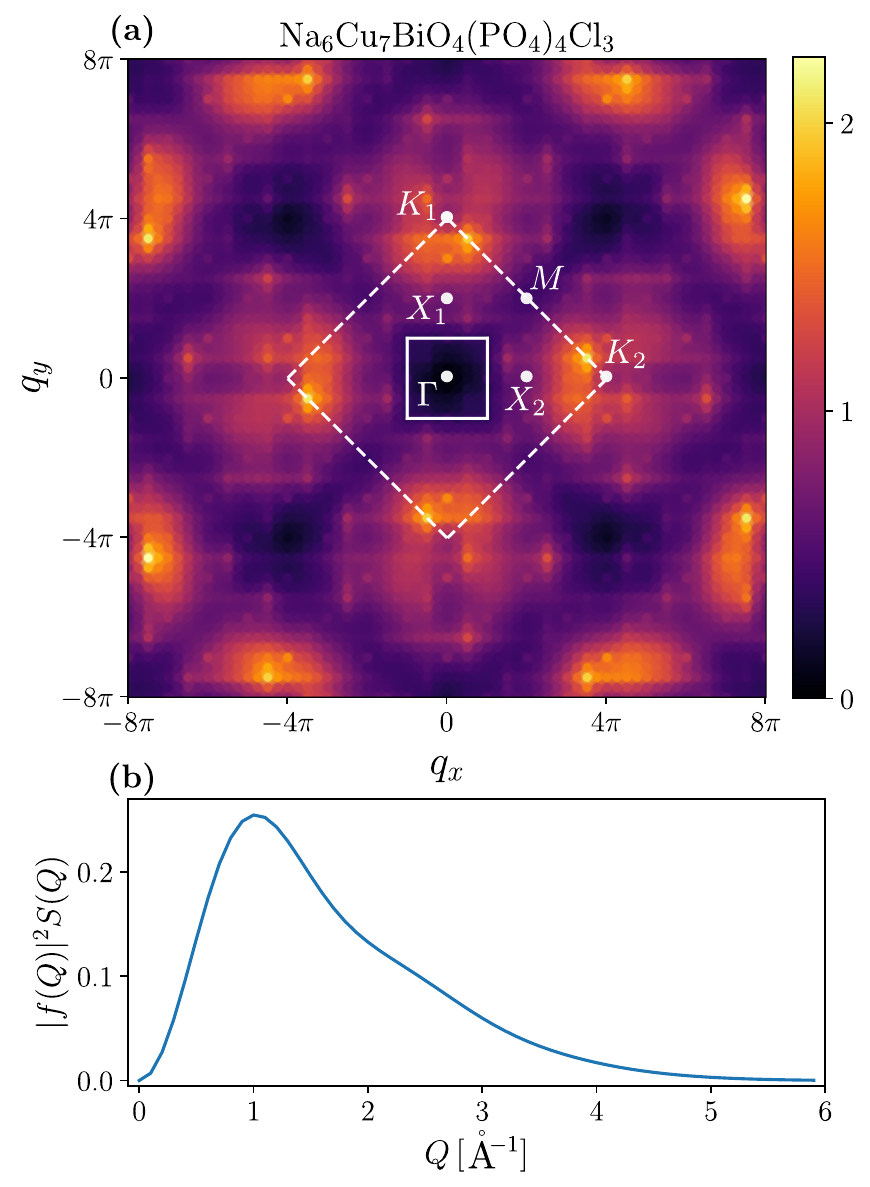}
\caption{ (a) Static (equal-time) structure factor from mVMC [Eq.~\eqref{eq:S_q}] obtained w.r.t.\ true crystal lattice site positions \cite{SM} obtained on a $8\times8\times7$ site cluster [Note: the ($S\mathbf{q}$) is not periodic, and the Brillouin zones and high-symmetry points of the ideal geometrical lattice are only drawn for illustrative purposes], (b) The corresponding powder average after accounting for the form factor. 
}\label{fig:equaltime}
\end{figure}

Interestingly, we notice that the inclusion of $J_{10}$ interactions in the Hamiltonian increases the frustration thereby enhancing the disordering tendency. This is reflected in an increase of the ground state energy per site [see Table~\ref{tab:mVMCenergies}] and decreasing correlations between Cu(1) and Cu(2) sites in favor of bonds containing a \Cuthree site (see Fig. S7 in Ref.~\onlinecite{SM}). In particular, the ratio of $J_{10}/J_{2}\sim0.34$ places the material in the vicinity of the high point of frustration (largest ground state energy) (Table S2 in Ref.~\onlinecite{SM}). In mVMC calculations, the effect of increasing $J_{10}$ is to induce the rotation and reflection symmetry breaking, and broadening of the maxima in S({\bf q}) leading to a more diffuse signal (see Fig. S4 in Ref.~\onlinecite{SM}). Thus, an important aspect of this result is that despite an appreciable magnetic coupling of the decorated \Cuthree ions with the square-kagome layers, it does not result in magnetic ordering. Quite the contrary, we show that the presence of these strongly correlated interlayer Cu$^{2+}$ ions aids the stabilization of a magnetically disordered ground state -- thus settling the question raised by specific heat measurements~\cite{Liu2022}. It is interesting to note that in a related square-kagome material nabokoite KCu$_7$TeO$_4$(SO$_4$)$_5$Cl which similarly features decorating Cu sites, signatures of long-range ordering have recently been reported~\cite{Markina2022}.

The symmetry breaking manifests itself in the static spin structure factor which is likewise $C_{2}$ symmetric [see Fig.~\ref{fig:equaltime}(a)] as obtained from mVMC. The maxima are located at $\mathbf{q}=\pm 2\pi(3.75, 2.25)$ with the follow-up maxima at $\mathbf{q}=\pm 2\pi(1.75, 0.25)$. The powder average structure factor is presented in Fig.~\ref{fig:equaltime}(b) in order to facilitate comparison with potential neutron scattering studies.

\begin{figure}
  \includegraphics[width = 1.0\textwidth]{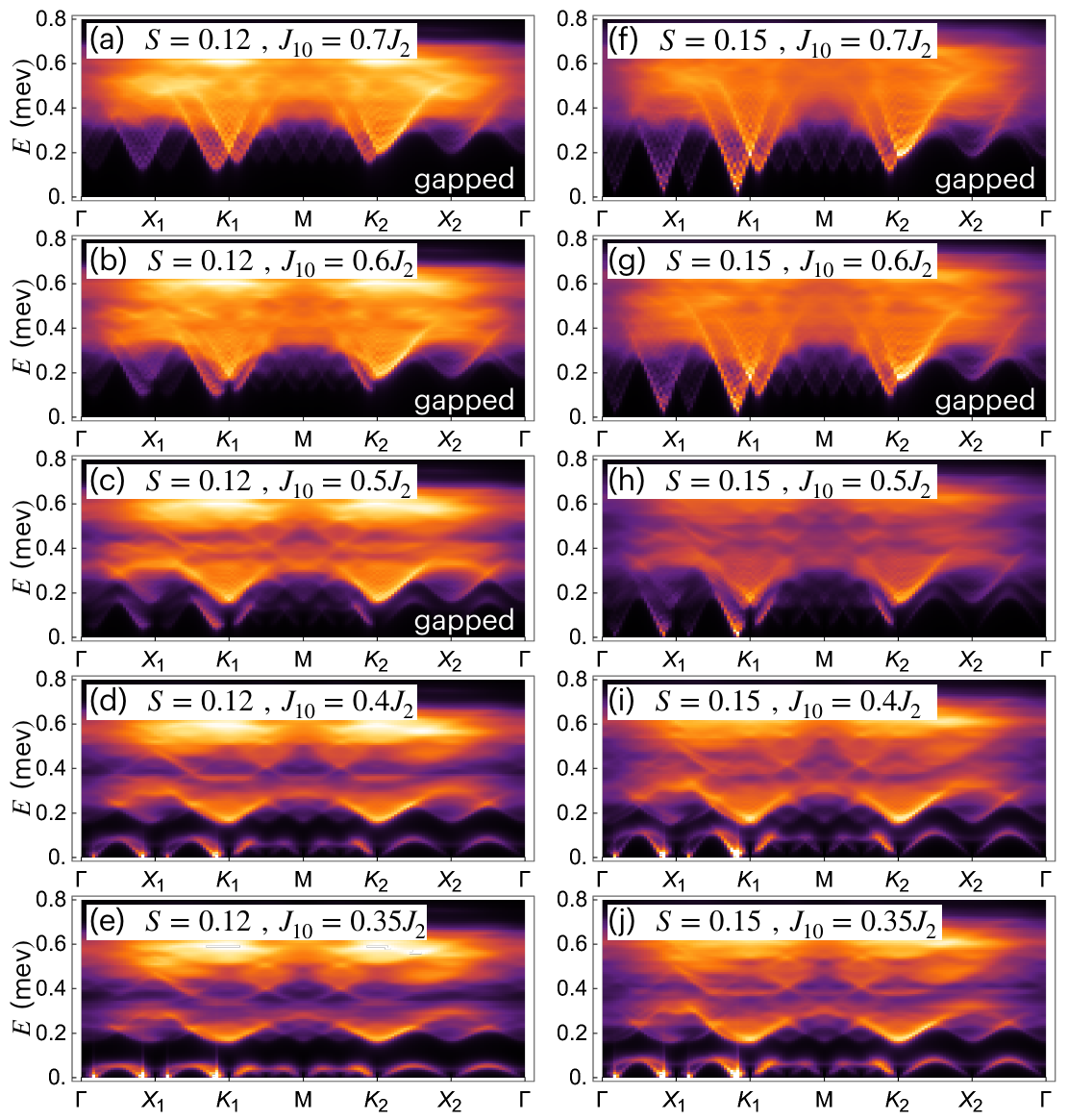}
  \caption{Dynamical structure factors as function of $S$ and $J_{10}$. The other parameters are as given in the caption of Fig~\ref{fig:exchange}. As $J_{10}$ increases and/or $S$ decreases, a gap opens and a quantum paramagnet is stabilized. The Bose condensations appear at incommensurate q vectors. In the condensed state, the \Cuthree spins are ordered but the other spins in the square-kagome lattice remain very weakly ordered. This is reflected by the gap between the lower branch excitations and the continuum in the lower panels. }
  \label{fig:dynsbmft}
\end{figure}

The flexibility of the SBMFT method allows to efficiently compute the dynamical structure factor
\begin{eqnarray}
    S(\mathbf{q},\omega) &=&\frac{1}{N_s} \sum_{i,j} e^{i \bf{q} \cdot (\bf{r}_i-\bf{r}_j)} \int_{-\infty}^{\infty} dt e^{-i\omega t} \langle {\mathbf{\hat S}}_{i}(t){\mathbf{\hat S}}_{j}(0) \rangle,\nonumber \\ 
\end{eqnarray}
and to extract interesting magnon features with information on the Bose condensations of specific branches. Here, $N_s$ is the total number of sites given by $n_u \times 2 \times l \times l$, where $n_u$ is the number of sites per unit cell (here 14 in the presence of the \Cuthree atoms), and $l$ is the linear size of the system. This quantity can be compared with with neutron scattering experiments. Within this approach one can artificially tune $S$ to lower values in order to enhance quantum fluctuations~\cite{Lugan-2022,Messio-2012}. Thus, in Fig.~\ref{fig:dynsbmft}, we show the dynamical structure factor for two representative spin values $S = 0.12, 0.15$ for which a quantum paramagnetic ground state can be stabilized, and also for various values of Cu$(3)$ coupling, $J_{10}$, for a system size of $l=12$ with $4032$ spins. This figure displays several features: (i) decreasing $J_{10}$ or increasing $S$ favours Bose condensation of the \Cuthree atom spins, (ii) they appear at incommensurate values of the BZ, (iii) the rotational symmetry breaking is evident from asymmetric excitation spectra around the $K_1$ and $K_2$ points, see Fig.~\ref{fig:equaltime}(a) and (iv) the gap closes at the extracted parameters but a secondary gap between the lower branch and the continuum appears, reflecting ordered \Cuthree spins,
while the others remain weakly ordered. This allows us to reveal the proximity to a phase transition between a quantum paramagnetic state and its Bose condensate counterpart. 

As seen in the fermion approaches, the effect of projecting the wave function on exact physical states increases the quantum fluctuations and help the system to remain disordered even in the presence of the \Cuthree atoms. In the SBMFT, since magnetic orders are more competitive by construction, they are favoured at $S=1/2$. Thus, in order to reach the quantum paramagnet, one has to reduce the spin value.

We can see that the Bose condensation arises on the \Cuthree spins while the others on the square-kagome lattice remain disordered. This is reflected in Fig.~\ref{fig:dynsbmft} in a clear gap in the lower panels between the condensed branches and the excitations in the continuum [see Fig.~S11 of Supp. Mat.~\cite{SM} for real space spin correlation profiles]. As one can see, a clear magnetic order appears on the \Cuthree spins while the spins on the square-kagome lattice are disordered.

{\it Conclusions} We have determined a Heisenberg Hamiltonian for {\skag} with five significant antiferromagnetic exchange interactions. While triangle couplings in the square-kagome lattice dominate at about twice the size of the square couplings, the diagonals in the squares are the second largest interaction. The \Cuthree sites decorating the square kagome lattice in {\skag} turn out to be substantially coupled to the square sites. Our five numerical techiques all corroborate that the Hamiltonian has a nonmagnetic ground state, in agreement with the fact that experimentally, no order was found down to 50\,mK. We find the nature of this ground state to be a VBC which breaks translational symmetry. We predict static as well as dynamical structure factors to motivate studies of {\skag} with inelastic neutron scattering.

{\it Acknowledgments}. 
We thank A. Vasiliev and S. Streltsov for helpful discussions during the early stages of this project. 
N.\,A. is funded by the Swiss National Science Foundation, grant number: PP00P2{\_}176877. The mVMC simulations were supported by the RSF grant (project No.\,21-12-00237). 
Funding by the Deutsche Forschungsgemeinschaft (DFG, German Research Foundation) is acknowledged by F.\,F. through TRR 288 -- 422213477 (project A05), by N.\,N. and J.\,R. within Project-ID 277101999 CRC 183 (Project A04), J.\,Ri. through project RI 615/25-1, T.\,M. and R.\,T. through Project-ID 258499086-SFB 1170 and the Wurzburg-Dresden Cluster of Excellence on Complexity and Topology in Quantum Matter – ct.qmat Project-ID 390858490-EXC 2147.
The work of Y.I. was performed in part at the Aspen Center for Physics, which is supported by National Science Foundation grant PHY-2210452. The participation of Y.I. at the Aspen Center for Physics was supported by the Simons Foundation. The research of Y.I. was supported, in part, by the National Science Foundation under Grant No. NSF PHY-1748958. Y.I. acknowledges support from the ICTP through the Associates Programme and from the Simons Foundation through grant number 284558FY19, IIT Madras through the IoE program for establishing QuCenDiEM (Project No. SP22231244CPETWOQCDHOC), the International Centre for Theoretical Sciences (ICTS), Bengaluru, India during a visit for participating in the program “Frustrated Metals and Insulators” (Code: ICTS/frumi2022/9). Y.I. acknowledges the use of the computing resources at HPCE, IIT Madras.
F.\,F., A.\,R., N.\,N., R.\,T., J.\,R., and H.\,O,\,J. thank IIT Madras for funding a visiting research fellow position under the IoE program during which this collaboration was initiated, and the initial parts of the research work were performed. 
N.\,N. and J.\,R. acknowledge the use of the JUWELS cluster at the Forschungszentrum J\"ulich and the Noctua2 cluster at the Paderborn Center for Parallel Computing (PC$^2$).
N.\,A. acknowledges the usage of computing resources of the federal collective usage center ``Complex for simulation and data processing for mega-science facilities'' at NRC ``Kurchatov Institute''.
T.\,M. and R.\,T. gratefully acknowledge the Gauss Centre for Supercomputing e.\,V. for funding this project by providing computing time on the GCS Supercomputer SuperMUC at Leibniz Supercomputing Centre.  

\bibliography{squarekagome}

\end{document}


\title{Quantum paramagnetism in the decorated square-kagome antiferromagnet {\skag}\\-- Supplemental Material --}

\author{Nils Niggemann}
\thanks{These authors contributed equally.}
\affiliation{Dahlem Center for Complex Quantum Systems and Fachbereich Physik, Freie Universit\"at Berlin, 14195 Berlin, Germany}
\affiliation{Helmholtz-Zentrum für Materialien und Energie, Hahn-Meitner-Platz 1, 14109 Berlin, Germany}
\affiliation{Department of Physics and Quantum Centre of Excellence for Diamond and
Emergent Materials (QuCenDiEM), Indian Institute of Technology Madras, Chennai 600036, India}

\author{Nikita Astrakhantsev}
\thanks{These authors contributed equally.}
\affiliation{Department of Physics, University of Z\"urich, Winterthurerstrasse 190, CH-8057 Z\"urich, Switzerland}

\author{Arnaud Ralko}
\thanks{These authors contributed equally.}
\affiliation{Institut Néel, UPR2940, Université Grenoble Alpes, CNRS, Grenoble, FR-38042 France}
\affiliation{Department of Physics and Quantum Centre of Excellence for Diamond and
Emergent Materials (QuCenDiEM), Indian Institute of Technology Madras, Chennai 600036, India}

\author{Francesco Ferrari}
\thanks{These authors contributed equally.}
\affiliation{Institut f\"ur Theoretische Physik, Goethe Universit\"at Frankfurt, Max-von-Laue-Stra{\ss}e 1, 60438 Frankfurt am Main, Germany}
\affiliation{Department of Physics and Quantum Centre of Excellence for Diamond and
Emergent Materials (QuCenDiEM), Indian Institute of Technology Madras, Chennai 600036, India}

\author{Atanu Maity}
\affiliation{Department of Physics and Quantum Centre of Excellence for Diamond and
Emergent Materials (QuCenDiEM), Indian Institute of Technology Madras, Chennai 600036, India}

\author{Tobias M\"uller} 
\affiliation{Institut f\"{u}r Theoretische Physik und Astrophysik, 
  Julius-Maximilians-Universit\"at W\"{u}rzburg, Am Hubland, D-97074 W\"{u}rzburg, Germany}

\author{Johannes Richter}
\affiliation{Institut für Physik, Otto-von-Guericke-Universität Magdeburg, P.O. Box 4120, 39016 Magdeburg, Germany}
\affiliation{Max-Planck-Institut für Physik Komplexer Systeme, Nöthnitzer Straße 38, D-01187 Dresden, Germany}

\author{Ronny Thomale} 
\affiliation{Institut f\"{u}r Theoretische Physik und Astrophysik, 
  Julius-Maximilians-Universit\"at W\"{u}rzburg, Am Hubland, D-97074 W\"{u}rzburg, Germany}
\affiliation{Department of Physics and Quantum Centre of Excellence for Diamond and
Emergent Materials (QuCenDiEM), Indian Institute of Technology Madras, Chennai 600036, India}

\author{Titus Neupert}
\affiliation{Department of Physics, University of Z\"urich, Winterthurerstrasse 190, CH-8057 Z\"urich, Switzerland}

\author{Johannes Reuther}
\affiliation{Dahlem Center for Complex Quantum Systems and Fachbereich Physik, Freie Universit\"at Berlin, 14195 Berlin, Germany}
\affiliation{Helmholtz-Zentrum für Materialien und Energie, Hahn-Meitner-Platz 1, 14109 Berlin, Germany}
\affiliation{Department of Physics and Quantum Centre of Excellence for Diamond and
Emergent Materials (QuCenDiEM), Indian Institute of Technology Madras, Chennai 600036, India}

\author{Yasir Iqbal}
\affiliation{Department of Physics and Quantum Centre of Excellence for Diamond and
Emergent Materials (QuCenDiEM), Indian Institute of Technology Madras, Chennai 600036, India}

\author{Harald O. Jeschke}
\affiliation{Research Institute for Interdisciplinary Science, Okayama University, Okayama 700-8530, Japan}
\affiliation{Department of Physics and Quantum Centre of Excellence for Diamond and
Emergent Materials (QuCenDiEM), Indian Institute of Technology Madras, Chennai 600036, India}

\date{\today}

\maketitle


\newcommand{\beginsupplement}{%
        \setcounter{table}{0}
        \renewcommand{\thetable}{S\arabic{table}}%
        \setcounter{figure}{0}
        \renewcommand{\thefigure}{S\arabic{figure}}%
        \setcounter{equation}{0}
        \renewcommand{\theequation}{S\arabic{equation}}%
        \setcounter{page}{1}
     }

\renewcommand*{\citenumfont}[1]{S#1}
\renewcommand*{\bibnumfmt}[1]{[S#1]}
 
\beginsupplement

\maketitle

\section*{Additional DFT information}\label{sec:dft}

\begin{figure}[b]
	\centering
	\includegraphics[width=\columnwidth]{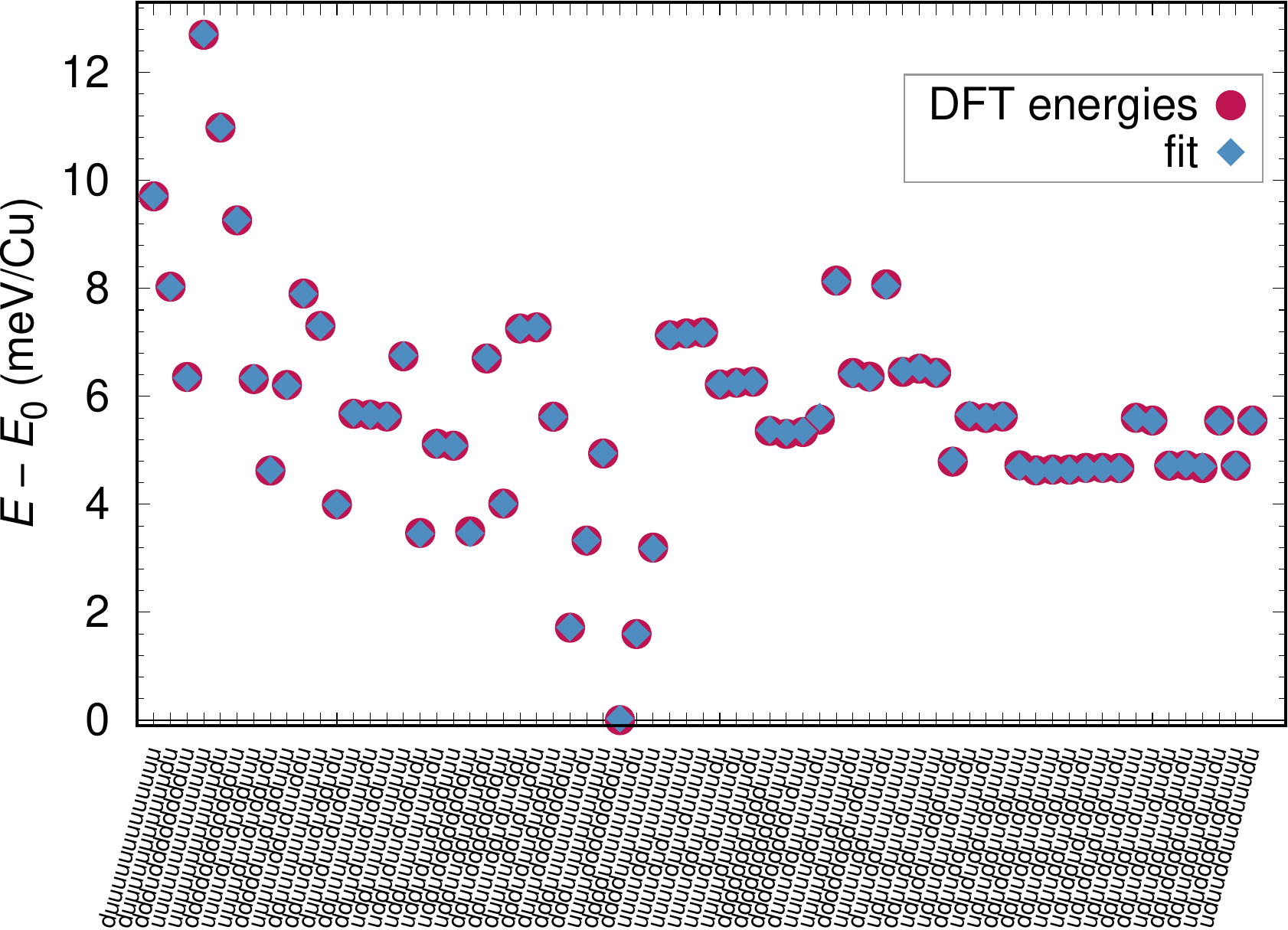}
	\caption{Quality of energy mapping in {\skag}. 68 calculated DFT energies are very precisely matched by the Heisenberg Hamiltonian with 13 exchange interactions.
	\label{fig:energies}}
\end{figure}

\begin{table*}[htb]
\centering
\caption{Exchange interactions of {\skag} obtained by DFT energy mapping as described in the main text. The line in bold face corresponds to the set of couplings that match the experimental Curie-Weiss temperature~\cite{Yakubovich2021}.}\label{tab:couplings}
{\sffamily \footnotesize
  \begin{tabular}{c|c|c|c|c|c|c|c|c|c|c|c|c|c|c}
$U$\,(eV) & $J_1$\,(K) & $J_2$\,(K) & $J_3$\,(K) & $J_4$\,(K) & $J_5$\,(K) & $J_6$\,(K) & $J_8$\,(K) & $J_9$\,(K) & $J_{10}$\,(K) & $J_{13}$\,(K) & $J_{14}$\,(K) & $J_{15}$\,(K) & $J_{20}$\,(K) & $\theta_{\rm CW}$\,(K)\\\hline
6.5 & 113.6(8) & 191.5(7) & 158.7(1.4) & 48.1(4) & 5.8(4) & 0.6(4) & -0.8(7) & 3.4(8) & 66.6(2) & -2.2(6) & 2.0(4) & 0.3(4) & -0.1(4) & -218.7 \\
{\bf 6.66} & {\bf 109.1(8)} & {\bf 186.2(7)} & {\bf 155.3(1.4)} & {\bf 46.9(4)} & {\bf 5.3(4)} & {\bf 0.6(4)} & {\bf -1.1(7)} & {\bf 3.2(8)} & {\bf 64.6(2)} & {\bf -2.2(6)} & {\bf 1.9(4)} & {\bf 0.3(4)} & {\bf 0.0(4)} & {\bf -212}\\
7 & 100.1(7) & 175.5(6) & 148.1(1.1) & 44.6(4) & 4.4(4) & 0.5(3) & -1.5(5) & 2.9(7) & 60.5(2) & -2.1(5) & 1.8(3) & 0.2(4) & 0.0(3) & -198.6\\
7.5 & 88.2(6) & 160.8(5) & 137.8(9) & 41.3(3) & 3.2(3) & 0.5(3) & -1.9(4) & 2.6(6) & 54.9(2) & -2.1(4) & 1.6(3) & 0.2(3) & 0.0(2) & -180.5\\
8 & 77.7(5) & 147.3(4) & 128.0(7) & 38.2(2) & 2.3(3) & 0.4(2) & -2.2(4) & 2.2(5) & 49.8(1) & -2.0(3) & 1.5(2) & 0.2(2) & 0.0(2) & -164.1\\
8.5 & 68.3(4) & 134.8(3) & 118.6(6) & 35.3(2) & 1.6(2) & 0.4(2) & -2.4(3) & 2.0(4) & 45.3(1) & -1.8(3) & 1.4(2) & 0.1(2) & 0.0(2) & -149.1 \\\hline 
$d$\,(\AA)&3.11405 & 3.27801 & 4.40394 & 5.0088 & 5.26566 & 5.33981 & 5.72746 & 5.76923 & 6.01815 & 6.55602 & 7.08351 & 7.71676 & 8.22822 &
\end{tabular}
}
\end{table*}

Density functional theory calculations were performed with the full potential local orbital (FPLO) basis~\cite{Koepernik1999} and a generalized gradient approximation (GGA) exchange correlation functional~\cite{Perdew1996}. Strong electronic correlations on Cu $3d$ orbitals are taken into account by a DFT+U method~\cite{Liechtenstein1995}. We show the full result of the energy mapping procedure for {\skag} in Table~\ref{tab:couplings}. For every considered value of the onsite interaction strength $U$, a total of 68 energies were calculated with $4\times4\times4$ $\mathbf{k}$-point meshes. The line in bold face is interpolated to yield the experimental Curie-Weiss temperature $\theta_{\rm CW}=-212$\,K~\cite{Yakubovich2021}. As discussed in the main text, only five of the thirteen determined exchange interactions are substantial and are considered in the many-body calculations for {\skag}. The only interlayer coupling $J_{15}$ we determined is small enough that {\skag} can be treated as a 2D system.

In Fig.~\ref{fig:energies}, we show that the energy mapping approach works very well for {\skag}. In $P1$ symmetry, all 14 Cu$^{2+}$ moments can be set independently, and for the 68 chosen spin configurations, the comparison between DFT energy and fit to the Heisenberg Hamiltonian is excellent.

\section*{Lattice Convention}\label{sec:latconv}
\begin{figure}
    \centering
    \includegraphics[width = 0.6\linewidth]{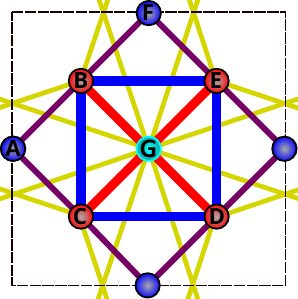}
    \caption{Idealized square-kagome unit cell. Sites A and F correspond to Cu(1) ions, sites B, C, D, E to Cu(2) ions and site G to the \Cuthree ion in the unit cell of {\skag}.}
    \label{fig:squagome2DUC}
\end{figure}

Each layer of the compound {\skag} is represented by a square-kagome lattice with an additional site at the center of each shuriken (square surrounded by four triangles). More precisely, we define the unit vectors
\begin{equation}
    \mathbf{a}_1 = (1,0) ,\quad \mathbf{a}_2 = (0,1),
\end{equation}
and a unit cell consisting of seven sites, labelled from $A$ to $G$ as shown in Fig.~\ref{fig:squagome2DUC}, whose relative coordinates are
\begin{align}\label{eqn:coord}
        \boldsymbol{\delta}_A &= (0,1/2),\ 
        \boldsymbol{\delta}_B = (1/4,3/4) \nonumber \\
        \boldsymbol{\delta}_C &= (1/4,1/4),\
        \boldsymbol{\delta}_D = (3/4,1/4) \nonumber \\
        \boldsymbol{\delta}_E &= (3/4,3/4),\ 
        \boldsymbol{\delta}_F = (1/2,1) \nonumber \\
        \boldsymbol{\delta}_G &= (2/4,2/4) 
\end{align}

In this convention, all sites lie at rational fractions of the basis vectors, which ensures a finite extended Brillouin zone over which the structure factor is periodic. The lattice for this compound features three types of symmetry inequivalent sites, namely those that lie on the vertices of each shuriken ($A,F$), the corners of each square ($B,C,D,E$) and the center sites $G$ [see Fig.~\ref{fig:squagome2DUC}].
Although, the geometry of the lattice does not alter the physics (once the couplings are fixed), for comparison with experiments, it is required to compute Fourier-transformed quantities with respect to the lattice of the original compound.
As most numerical methods do not explicitly employ real-space crystallographic coordinates but rather the more abstract graph of sites and respective bonds, this amounts to a choice in the post-processing of real-space data.
In the following, we outline this step in detail.
We define the Fourier transform of the spin susceptibility as
\begin{align}
    \chi(\mathbf{q}) = \frac{1}{N} \sum_{i,j} \chi(\mathbf{r}_i,\mathbf{r}_j) e^{i \mathbf{q}\cdot(\mathbf{r}_i - \mathbf{r}_j)}. \label{eq:chi3D}
\end{align}
Here, the vectors $\mathbf{q} , \mathbf{r}_i$ are three-dimensional, accounting for the true crystallographic coordinates, and allowing for distortions outside of the square-kagome planes which occur in the crystal structure. We must therefore map each pair of sites of the true 3D lattice to a unique site in the idealized lattice.

Upon noting that the layers are decoupled (as we neglect the tiny $J_{15}$ exchange interaction) and correlations between two sites from different layers are zero, we can immediately interpret the summation to go over sites in a single layer only. 

\begin{figure}
    \centering
    \includegraphics[width = 0.99\linewidth]{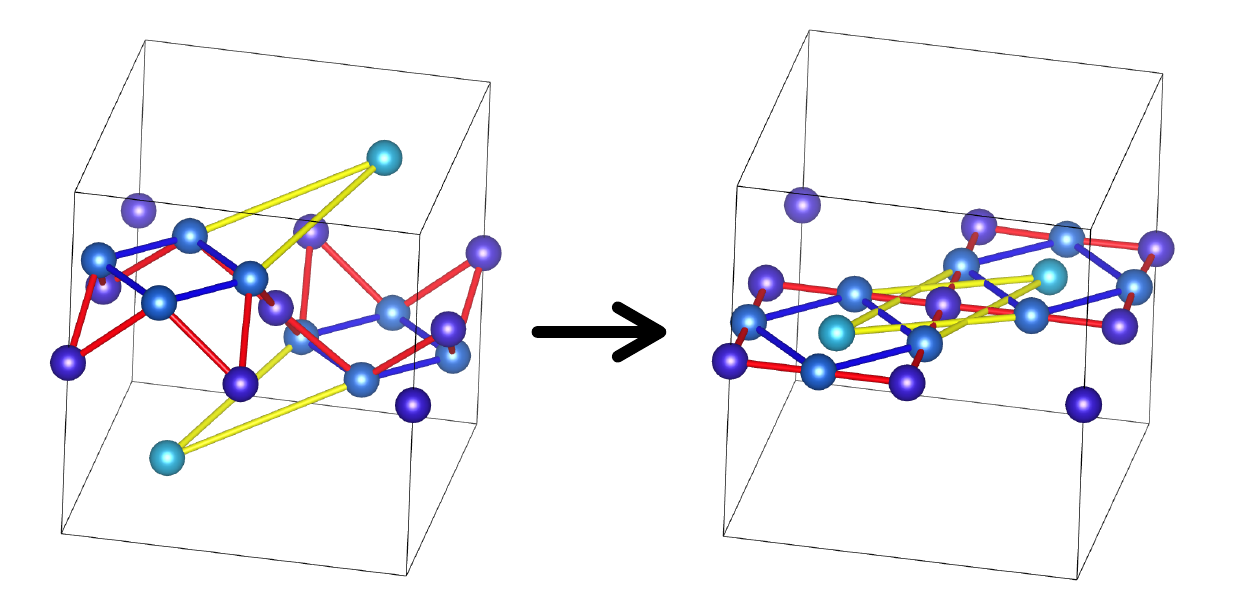}
    \caption{Mapping from 3D lattice to 2D lattice. To compute Eq.~\eqref{eq:chi3D}, sites are mapped to the idealized lattice convention. Visualization done using VESTA \cite{momma2011vesta}.}
    \label{fig:3DMapping}
\end{figure}

The mapping we need to perform can then be written as $f(n_1,n_2,0,n_b) = (n_1',n_2',n_b')$, where we define $\mathbf{r}_i = n_1 \mathbf{a}_1 +  n_2 \mathbf{a}_2 +  \mathbf{b}_{n_b}$.
Since the unit cell of the three-dimensional lattice can be enlarged compared to the idealized one, it may be cumbersome to find this mapping directly. A convenient strategy is to work in Cartesian coordinates, mapping $\mathbf{r}_i$ to a $2D$ point $(x,y)$ which lies on the idealized lattice. Then, the Cartesian coordinates $(x,y)$ can be transformed back into either a lattice position $(n_1,n_2,n_b)$, or a site index $i$. This is shown in Fig.~\ref{fig:3DMapping}.

To compare with powder samples, we compute the powder averaged structure factor
\begin{equation}
    S(Q) \equiv \frac{1}{4 \pi} \int d \Omega S(\mathbf{q}), \quad Q = |\mathbf{q}| \label{eq:powderavg}
\end{equation}
and multiply by the form factor to account for nuclear scattering on Cu$^{2+}$ ions \cite{Ghosh2023}
\begin{align}
 f(Q) &= A e^{-a  \frac{Q^2}{4\pi}} + B e^{-b  \frac{Q^2}{4\pi}} + C_1 e^{-c  \frac{Q^2}{4\pi}} + D_1\\
A &= 0.0232,\ B = 0.4023\nonumber \\
C_1 &= 0.5882,\ D_1 = -0.0137\nonumber \\
a &= 34.969,\ b = 11.564,\ c = 3.843\nonumber.
\end{align}

\section*{Methods}

\begin{figure*}[t!]
  \includegraphics[width = \linewidth]{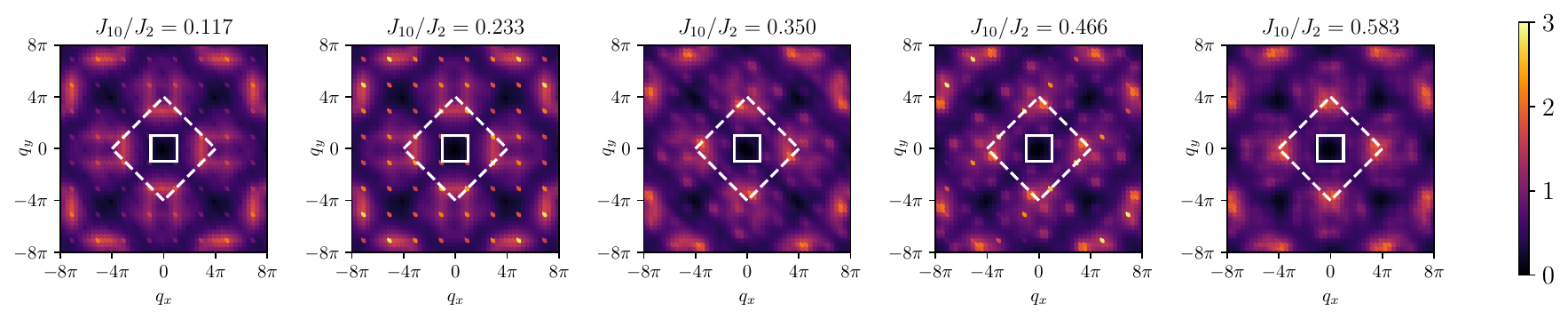}
  \caption{Momentum-resolved equal-time spin structure factor calculated from mVMC on the $6 \times 6 \times 6$ lattice for different values of $J_{10} / J_2$. The Fourier transform is performed with respect to the ideal lattice coordinates of Eq.~\eqref{eqn:coord}. The wave functions were obtained with the symmetrization technique, enforcing the trivial representation. The solid and dashed lines denote the first and the extended Brillouin zones, respectively. The corresponding energies of these wave functions are given in Table~\ref{tab:j10_scan_table}.}
  \label{fig:j10_scan}
\end{figure*}

\subsection{Many-variable variational Monte Carlo (mVMC)}
\label{sec:mVMC}
In this work, we employ the many-variable variational Monte Carlo implementation presented in Refs.\,\cite{misawa2019mvmc,doi:10.1143/JPSJ.77.114701}. To apply mVMC, spin operators are first mapped to pseudo-fermionic bilinears through the Abrikosov representation
\begin{equation}\label{eq:Sabrikosov_fermions}
\hat{\mathbf{S}}_i=\frac{1}{2}\sum_{\alpha,\beta=\downarrow,\uparrow} \hat{c}_{i,\alpha}^{\dagger} \boldsymbol{\sigma}_{\alpha,\beta} \hat{c}_{i,\beta}\,.
\end{equation}

Inspired by Anderson's resonating valence-bond wave function, the mVMC ansatz has the form
\begin{equation}
    |\phi_{\mbox{\footnotesize pair}} \rangle = \hat{\mathcal{P}}^{\infty}_{\mbox{\footnotesize G}} \exp \left(\sum\limits_{i, j} f_{i,j} \hat c^{\dagger}_{i, \uparrow} \hat c^{\dagger}_{j \downarrow} \right) |0 \rangle,
\end{equation}
where a fermionic wave function is constrained to a space of singly occupied sites by means of the Gutzwiller projector $\hat{\mathcal{P}}^{\infty}_{\mbox{\footnotesize G}}=\prod_i (n_{i,\uparrow}-n_{i,\downarrow})^2$, where $n_{i,\sigma}=\hat c^{\dagger}_{i, \sigma}\hat c_{i, \sigma}$. The wave-function value $\langle \boldsymbol{\sigma} |\phi_{\mbox{\footnotesize pair}} \rangle $ for a specific spin configuration $|\boldsymbol{\sigma}\rangle$ is evaluated using the Slater determinant of the matrix with elements $f_{i,j}$. The parameters $f_{i,j}$ are optimized using the stochastic reconfiguration optimization technique~\cite{sorella_green_1998,becca_quantum_2017,carleo2017solving}.

\begin{table}[b]
    \centering
    \begin{tabular}{ccccccc}
    \hline\hline
$J_{10} / J_2$ & $0.0$ & $0.117$ & $0.223$ & $0.350$ & $0.466$ & $0.583$ \\\hline
$E / J_2$ & $-0.4128$ & $-0.4128$ & $-0.4080$ & $-0.4305$ & $-0.4436$ & $-0.4687$  \\ \hline\hline

    \end{tabular}
    \caption{Ground state energy $E / J_2$ per site obtained on the $6 \times 6$ lattice within mVMC for different ratios of $J_{10} / J_2$. The statistical error is kept around $\delta E / J_2 = 10^{-4}$.}
    \label{tab:j10_scan_table}
\end{table}

We may force a wave function to transform as a specific irreducible representation of the symmetry group. To this end, we apply its generators until the symmetry orbit is exhausted
\begin{equation}
 |\Psi_\xi \rangle = \hat P |\Psi \rangle =  \sum\limits_n \xi^n \hat G^n |\Psi \rangle,
    \label{eq:q_projection}
\end{equation}
where $\xi$ is the desired projection quantum number and $|\Psi_\xi \rangle$ the resulting symmetrized state. In the \texttt{mVMC} package, the projection onto the total spin $S$ is performed by superposing the $SU(2)$--rotated wave functions~\cite{doi:10.1143/JPSJ.77.114701}. 

In Fig.~\ref{fig:j10_scan}, we show the equal-time spin structure factor for different values of $J_{10} / J_2$ measured within mVMC on the $6 \times 6$ lattice. The corresponding wave function energies are given in Table~\ref{tab:j10_scan_table}. One may see that the frustration grows until $J_{10} / J_2 \approx 0.3$, which is manifested by growing energy and progressively diffuse character of the structure factor, and then rapidly decays with further increase in $J_{10}/J_{2}$. 

\subsection{Variational Monte Carlo (VMC)} \label{sec:VMC}

Analogously to mVMC, the fermionic variational Monte Carlo calculations of this work make use of a Gutzwiller-projected wave function to approximate the ground state of the spin Hamiltonian by minimization of the variational energy. The VMC trial state reads
\begin{equation}
|\Psi_0\rangle=\hat{\mathcal{J}}\hat{\mathcal{P}}^{\infty}_{\mbox{\footnotesize G}} |\Phi_0\rangle,
\end{equation}
where ${\hat{\mathcal{P}}}^{\infty}_{\mbox{\footnotesize G}}$ is the Gutzwiller projector, $|\Phi_0\rangle$ is a fermionic Slater determinant [in the Abrikosov fermion representation of spins Eq.~\eqref{eq:Sabrikosov_fermions}] and $\hat{\mathcal{J}}$ is a spin-spin Jastrow factor. The definition of the fermionic state $|\Phi_0\rangle$ is done by the introduction of an auxiliary fermionic Hamiltonian
\begin{equation}\label{eqn:mf-nonmag}
\hat{\mathcal{H}}_{\rm 0}=\sum_{i,j,\alpha}\chi_{ij}\hat{c}_{i,\alpha}^{\dagger}\hat{c}_{j,\alpha},
\end{equation}
whose hopping amplitudes, $\chi_{ij}$, play the role of variational parameters. The presence of the spin-spin Jastrow factor
\begin{equation}
\hat{\mathcal{J}}=\exp\left(\sum_{i,j} v_{i,j} \hat{S}^z_i \hat{S}^z_j \right)
\end{equation}
can enhance/suppress spin-spin correlations between lattice sites. We consider a long-range Jastrow factor with pseudopotential parameters that depend on the distance between sites, i.e., $v_{i,j}=v(\|\mathbf{r}_i-\mathbf{r}_j\|)$. The variational parameters defining $|\Psi_0\rangle$ are optimized by means of the stochastic reconfiguration method~\cite{sorella_green_1998,becca_quantum_2017}. The VMC calculations are performed on finite-size lattices with periodic boundary conditions. Specifically, we employ two kinds of fully-symmetric clusters, one defined by the translation vectors $\mathbf{T}_1=L\mathbf{a}_1$ and $\mathbf{T}_2=L\mathbf{a}_2$ (containing $N_s=7L^2$ sites), and the other defined by the vectors  $\mathbf{T}_1=L(\mathbf{a}_1+\mathbf{a}_2)$ and $\mathbf{T}_2=L(\mathbf{a}_1-\mathbf{a}_2)$ (containing $N_s=14L^2$ sites). The biggest lattice employed in our VMC calculations contains $N_s=700$ spins.

\begin{figure}
  \includegraphics[width = \linewidth]{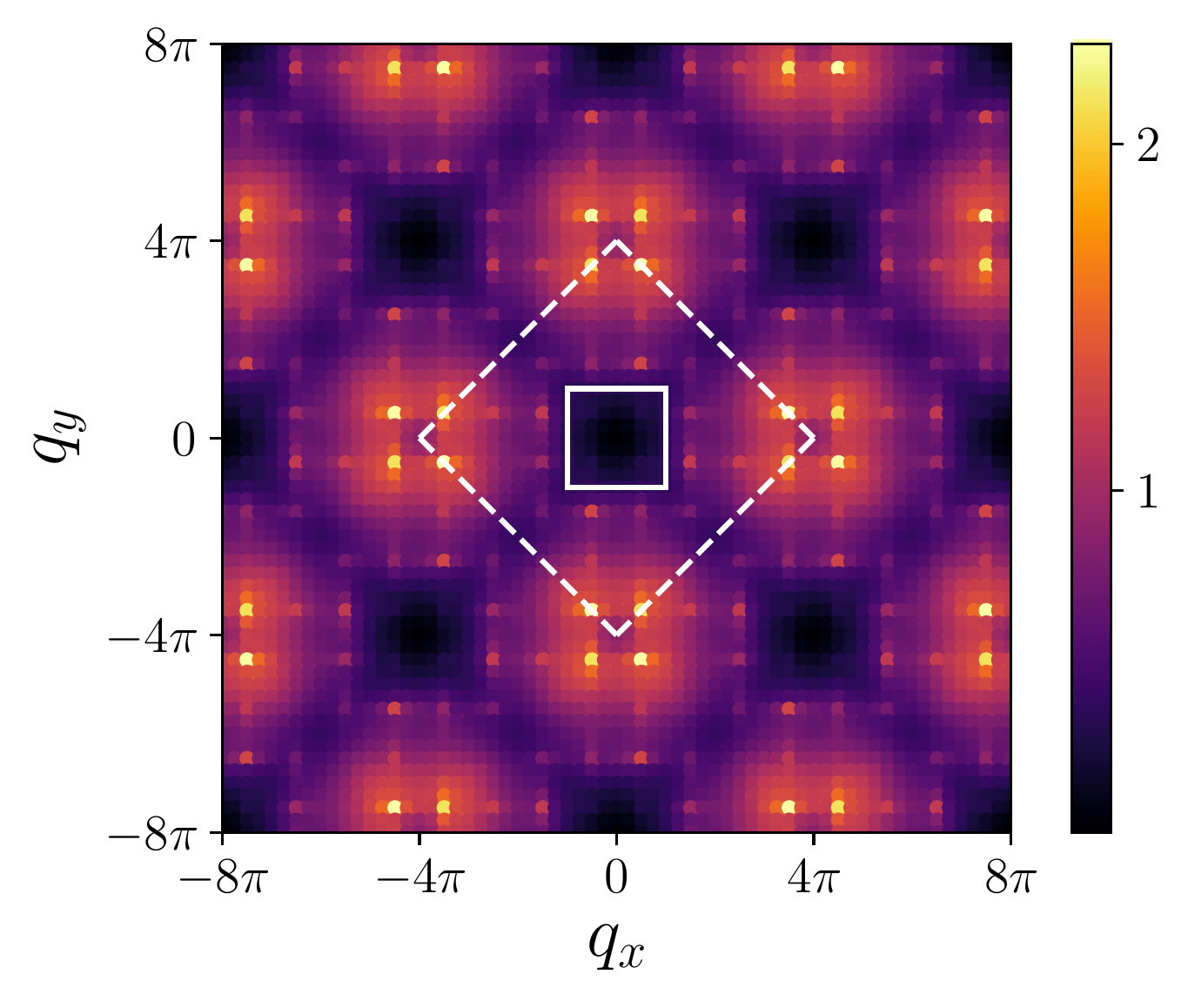}
  \caption{Equal-time spin structure factor $S(\mathbf{q})$ from VMC for the Hamiltonian of {\skag}. The results are obtained on a cluster with $N_s=448$ sites, defined by the translation vectors $\mathbf{T}_1=8\mathbf{a}_1$ and $\mathbf{T}_2=8\mathbf{a}_2$, and plotted within the reciprocal space of the idealized lattice. The solid
and dashed lines denote the first and the extended Brillouin zones, respectively.}
  \label{fig:vmc_sq}
\end{figure}

The best variational energy is obtained by a parametrization of the auxiliary Hamiltonian in which the translational symmetry is broken. In particular, in agreement with mVMC observations, the optimal hopping pattern requires a doubling of the unit cell along $\mathbf{a}_1$ and $\mathbf{a}_2$ ($2\times 2$ periodicity). Concerning point group symmetries, the VMC wave function turns out to preserve the $C_4$ rotational symmetry around the center of the $2\times 2$ supercell. The small $C_4$ to $C_2$ symmetry breaking found by mVMC calculations is not captured by the VMC variational ansatz. The spin structure factor $S(\mathbf{q})$ corresponding to the best variational state for the Heisenberg Hamiltonian of {\skag} is shown in Fig.~\ref{fig:vmc_sq}. The absolute maxima of $S(\mathbf{q})$ are located at $\mathbf{q}=(7\pi/2,\pi/2)$ and $\mathbf{q}=(\pi/2,7\pi/2)$ (and symmetry related points). Compared to mVMC results, we find a larger intensity of the peaks of the structure factor, which can be ascribed to the presence of the Jastrow factor that enhances antiferromagnetic correlations.

\subsection{PMFRG}
The PMFRG formalism expresses spin operators by three different flavors ($x,y,z$) of Majorana fermions
\begin{equation}
  S^x_i = -i \eta^y_i \eta^z_i \text{,} \qquad  S^y_i = -i \eta^z_i \eta^x_i \text{,} \qquad  S^z_i = -i \eta^x_i \eta^y_i \text{.} \label{eq:MajoranaRep}
\end{equation}
This representation has the advantage that no unphysical states are introduced, allowing for quantitatively correct predictions at finite temperature \cite{Niggemann2021,Niggemann2022}.
One of the key advantages of PMFRG is its high momentum space resolution which allows for an easy detection of incommensurate order. The only restriction is a cut-off of all correlations beyond a numerically chosen maximum distance $L$. In a paramagnet, correlation lengths are typically small and effectively zero beyond a characteristic correlation length $\xi$, which makes this approximation virtually exact. When the correlation length diverges, for instance at the critical point of a phase transition, a finite-size scaling analysis may be performed to give accurate estimates of the critical temperature \cite{Niggemann2022}.

\begin{figure}[t]
  \includegraphics[width = 1.0\columnwidth]{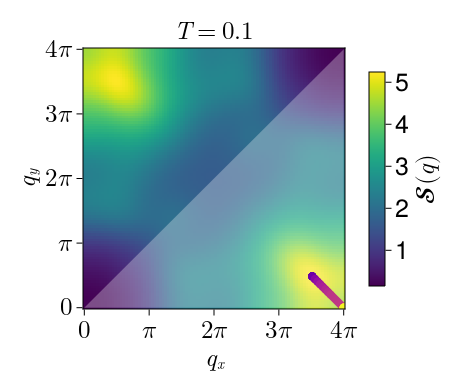}
  \caption{Evolution of the maxima of the equal-time spin structure factor as a function of temperature obtained from PMFRG. The color of the line corresponds to the temperature, i.e., the yellow part corresponds to the highest simulated temperature ($T=2.5 J_2$, while the blue part indicates the lowest temperatures. The background shows the low temperature structure factor as a reference.}
  \label{fig:pmfrg_maxima}
\end{figure}

\begin{figure}[b]
    \centering
    \includegraphics[width = \linewidth]{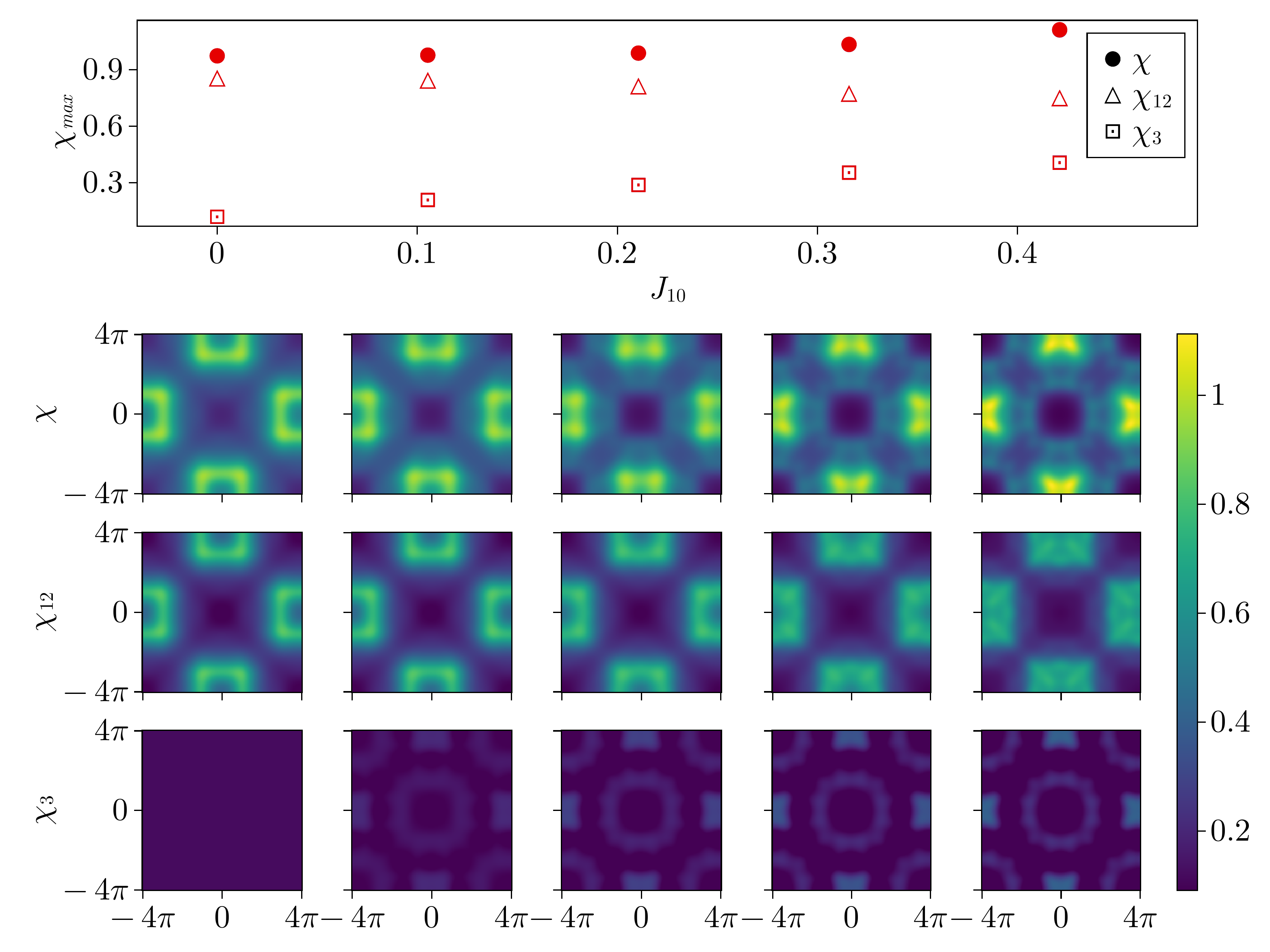}
    \caption{From PMFRG, we show the evolution of the sublattice dependent susceptibilities defined in Eq.~\eqref{eq:chi123} as a function of $J_{10}$. The top panel shows the respective maxima in momentum space, while the heat maps show the full momentum dependence. Note that points are shown for three different maximal correlation lengths of $L = 14,16,18$ nearest neighbor bonds, which all coincide, indicating absence of finite size effects.}
    \label{fig:PMFRGJ10Sweep}
\end{figure}

Figure~\ref{fig:pmfrg_maxima} shows the equal-time spin structure factor at low temperature, featuring soft maxima at wave-vectors $(q_x,q_y) \sim (3 \pi,\pi/2)$.
We find the position of these maxima to be temperature dependent. At high temperatures $T\gg J_2$, the maxima reside at $(4\pi,\pi)$ and symmetry related points, while the position of the peak shifts as temperature is lowered.
In addition to the spin susceptibility $\chi$ defined in Eq.~\eqref{eq:chi3D}, one may also define a sublattice-resolved susceptibility $\chi_{\alpha \beta},\ \alpha, \beta = 1,2,\dots N_\textrm{UC}$, where $N_\textrm{UC} = 7$ such that $\chi(\mathbf{k}) = \frac{1}{N_\textrm{UC}}\sum_{\alpha \beta}{\chi_{\alpha \beta}(\mathbf{k}) }$.
To study the influence of the \Cuthree sites, Fig.~\ref{fig:PMFRGJ10Sweep} further displays the susceptibilities 
\begin{align}
    \chi_{12}(\mathbf{k}) &\equiv \frac{1}{N_\textrm{UC}}\sum_{\alpha \beta \neq \Cuthree}\chi_{\alpha \beta}(\mathbf{k})\nonumber\\
    \chi_{3}(\mathbf{k}) &\equiv \chi(\mathbf{k}) - \chi_{12}(\mathbf{k}). \label{eq:chi123}
\end{align}

Here, $\chi_{12}$ quantifies the correlations between all non-\Cuthree sites, and $ \chi_{3}$ the contribution to the susceptibility upon adding the \Cuthree site.
The correlations between the Cu(1) and Cu(2) sites decrease upon strengthening the $J_{10}$ bond while maxima of intensity become diffuse, indicating increased frustration. To see this, note that the conservation of spin magnitude dictates a sum rule within the extended BZ: $\sum_i \mathbf{\hat S}_i^2 = \sum_{\mathbf q} \langle \mathbf{\hat S}(\mathbf{-q})\mathbf{\hat S}(\mathbf{q} \rangle) = \mathrm{N}\frac{3}{4}$. As a consequence, the susceptibility $\chi_{\alpha \beta}(\mathbf{q})$ for non-interacting spins must be a non-zero constant (see, e.g., $\chi_3$ in Fig.~\ref{fig:PMFRGJ10Sweep} at $J_{10} = 0$). Switching on the couplings between sites, leaves the sum rule intact for each sublattice but intensities may shift in reciprocal space and develop features. Therefore, the observed decrease in the maximum of $\chi_{12}$ indicates a more uniform distribution of different ordering wavevectors $\mathbf{q}$ within the system \textendash a signature of the effects of frustration.  

While the correlations between \Cuthree sites and the rest of the system naturally grow as they are increasingly coupled, ordering tendencies are absent as visible from the independence of our data on the maximal correlation length $L$ above the accessible temperature range of $T \gtrsim 0.3 J_{2}$. 

\begin{figure}
    \centering
    \includegraphics[width = \linewidth]{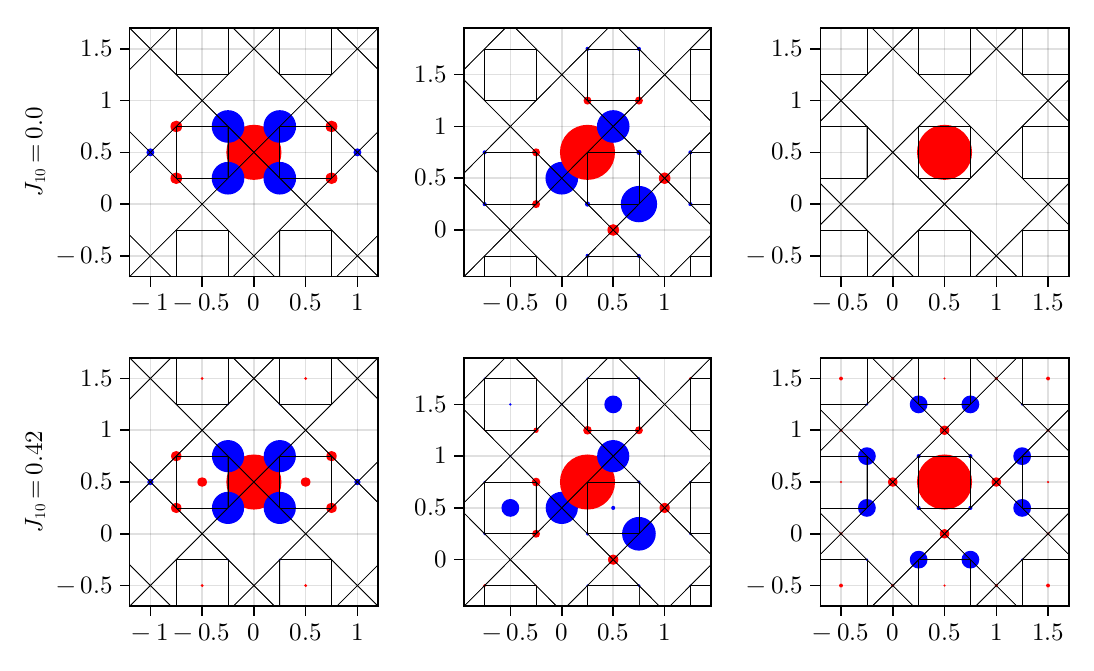}
    \caption{Real space correlations $\langle \hat{\mathbf{S}}_i \cdot \hat{\mathbf{S}}_j \rangle$ computed from PMFRG by taking Cu(1) (left), Cu(2) (middle), and \Cuthree (right) as reference sites $i$. The radii of the circle indicates the strength of correlations, while the color red (blue) indicates a positive (negative) sign of the correlations.}
    \label{fig:RealSpaceCorr_Cu}
\end{figure}

In Fig.~\ref{fig:RealSpaceCorr_Cu}, we also show the pattern of equal time spin-spin correlations in real space for the system with $J_{10}=0$ and $J_{10} \approx 0.42 J_2$. It is visible that adding the additional \Cuthree site does not lead to increased correlations between the other sites and, in fact, slightly decreases them. Overall, we find good agreement to mVMC in Fig.~2 of the main text.

\begin{figure}
  \includegraphics[width = \linewidth]{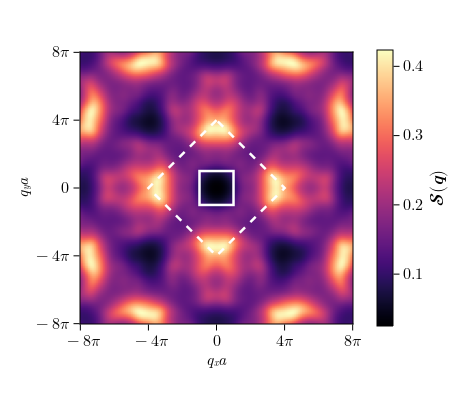}
  \caption{Equal time structure factor $S(q,t=0)$ at $T=0.3J_2$ for a cut at $q_z = 0$ computed with respect to the true, three dimensional crystallographic unit cell. Shown is the first BZ (solid), and the approximate extended BZ (dashed).}
  \label{fig:pmfrg_realFT}
\end{figure}

\begin{figure}
  \includegraphics[width = \linewidth]{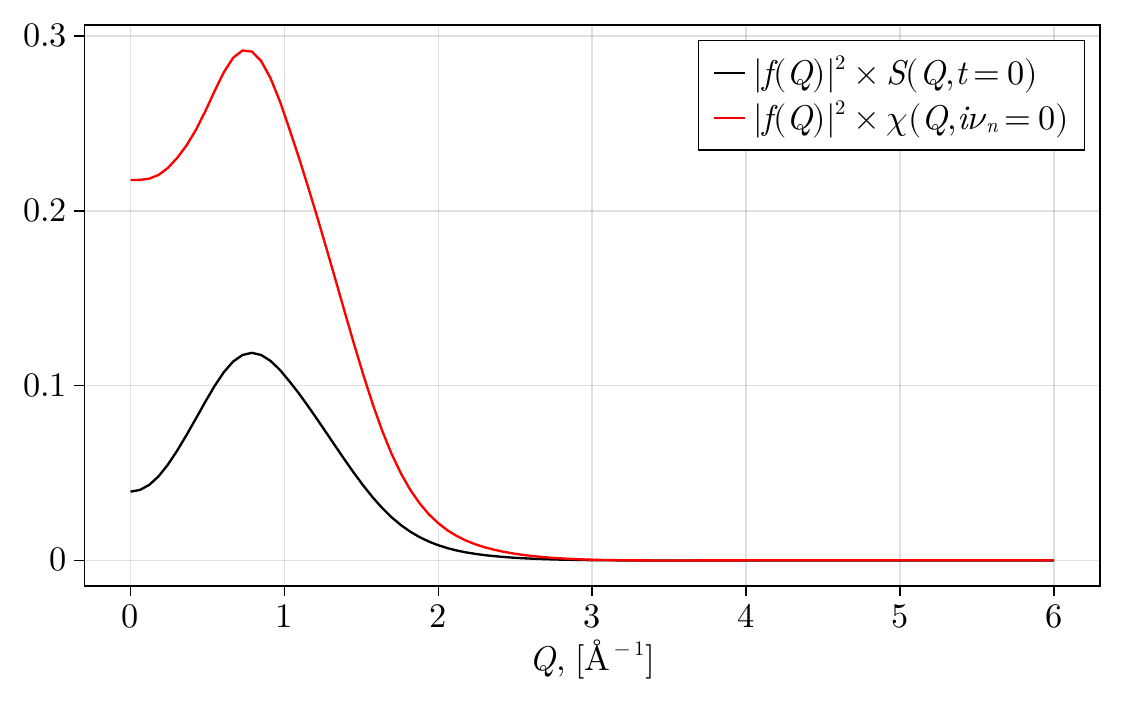}
  \caption{Powder average of the equal time structure factor $S(Q,t=0)$ and the static magnetic suscepibility $\chi(Q,i\nu_n=0)$ at $T=0.1J_2$ obtainde from PMFRG. As in Fig.~\ref{fig:pmfrg_realFT}, the Fourier transform is performed with respect to the actual crystal lattice site coordinates.}
  \label{fig:pmfrg_powderaverage}
\end{figure}

For comparison with experimental data, we compute the spin structure factor with respect to the actual crystal lattice site coordinates, i.e., allowing for atoms to be positioned outside of a pure 2D layer. Figure~\ref{fig:pmfrg_realFT} shows the corresponding momentum resolved equal-time spin structure factor projected onto the $q_x-q_y$ plane taking $q_z=0$. We note that due to the irrational coordinates of the atoms within the unit cell, the structure factor is no longer periodic within any extended Brillouin zone. The structure factor permits points of high intensity away from $q_z=0$, although they will naturally be broad, as length of correlations is limited in $z$ direction due to the layered structure of the material. If we restrict ourselves to a finite box in momentum space between $|q_{x,y,z}| < 8 \pi /a$, the maxima are positioned at incommensurate positions $\mathbf{k} = \left(0.61, 6.95, 0\right)\pi/a$.
For experimental comparison, in Fig.~\ref{fig:pmfrg_powderaverage}, we further provide the powder averaged structure factor \cref{eq:powderavg}.
\section*{Schwinger Boson mean field theory}

As in the VMC section, it is possible to make a parton construction of the spin by introducing Schwinger bosons instead of Abrikosov fermions. 
We thus consider the decoupling:
\begin{equation}\label{eq:Sabrikosov}
\hat{\mathbf{S}}_i=\frac{1}{2}\sum_{\alpha,\beta=\downarrow,\uparrow} \hat{b}_{i,\alpha}^{\dagger} \boldsymbol{\sigma}_{\alpha,\beta} \hat{b}_{i,\beta},
\end{equation}
where $\hat{b}^{(+)}$ are now bosonic operators. 
The advantage of dealing with bosons as opposed to fermions is the possibility of having Bose condensation and thus to easily access quantum magnetic orders as well as ($\mathds{Z}_2$) quantum spin liquids, which are both treated on an equal footing.
Here, we recall the main lines of the approach noting that more details can be found in \cite{Auerbach1998,Halimeh2016,Schaffer2017,Lugan-2019,Lugan-2022}.
In this approach, it is possible to write Heisenberg terms as function of two SU(2) invariant operators $\hat{A}$ and $\hat{B}$
\begin{eqnarray}
    \hat{A}_{ij} &=& \frac{1}{2} \left[ \hat{b}_{i\uparrow}^{}\hat{b}_{j\downarrow}^{} - \hat{b}_{i\downarrow}^{}\hat{b}_{j\uparrow}^{} \right] \nonumber \\
    \hat{B}_{ij} &=& \frac{1}{2} \left[ \hat{b}_{i\uparrow}^{+}\hat{b}_{j\uparrow}^{} + \hat{b}_{i\downarrow}^{+}\hat{b}_{j\downarrow}^{} \right] \nonumber 
\end{eqnarray}
as
\begin{eqnarray}
    \hat{\mathbf{S}}_i \hat{\mathbf{S}}_j &=& : \hat{B}_{ij}^+ \hat{B}_{ij} : -  \hat{A}_{ij}^+  \hat{A}_{ij},
\end{eqnarray}
where $::$ denotes the normal ordering.
At the mean field level, the Hamiltonian then reads 
\begin{eqnarray}
    \cal{H}_{\text{SB}} &=& = \sum_{i,j} J_{ij} \left[ \hat{B}_{ij}^+ {B}_{ij} +\hat{B}_{ij} {B}_{ij}^* - \hat{A}_{ij}^+  {A}_{ij} - \hat{A}_{ij} {A}_{ij}^*\right] \nonumber \\
    &-& \sum_{i,j} J_{ij} \left[ |B_{ij}|^2 - |A_{ij}|^2 \right] + \sum_i \lambda_i (\hat{n}_i - 2 S),
\end{eqnarray}
with the mean field parameters $A_{ij} = \langle \phi_0 | \hat{A}_{ij} | \phi_0 \rangle$ and $B_{ij} = \langle \phi_0 | \hat{B}_{ij} | \phi_0 \rangle$ computed in $|\phi_0 \rangle$ \textendash~the boson vacuum at $T=0$ for each pair of interacting spins $(i\to j)$. Because the Hilbert space is enlarged by the mapping, it is necessary to fulfill the constraint $\hat{n}_i = \hat{b}_{i \uparrow}^+ \hat{b}_{i \uparrow} + \hat{b}_{i \downarrow}^+ \hat{b}_{i \downarrow} = 2S$ for a spin $S$. Thus, we have also introduced Lagrange multipliers $\lambda_i$ to account for this on average. Another advantage of the method is that $S$ can be treated as an external parameter and by reducing it, it is possible to enhance quantum fluctuations. This is particularly interesting if one wants to focus on phase transitions between a magnetically ordered state and its quantum spin liquid parent. Also, the flexibility of the method allows to compute the dynamical structure factor
\begin{eqnarray}
    S(\mathbf{q},\omega) &=&\frac{1}{n_s} \sum_{i,j} e^{i \bf{q} \cdot (\bf{r}_i-\bf{r}_j)} \int_{-\infty}^{\infty} dt e^{-i\omega t} \langle \hat{\mathbf{S}}_{i}(t)\hat{\mathbf{S}}_{j}(0) \rangle,\nonumber \\ 
\end{eqnarray}
and to extract relevant magnon features and study Bose condensations of specific branches. Here, $n_s$ is the total number of sites given by $n_u \times 2 \times l \times l$, with $n_u$ the number of sites per unit-cell (here 14 in the presence of the \Cuthree atoms), and $l$ is the linear size of the system. This allows for comparison with neutron experiments. 

In Fig.~4 of the main text, we show the dynamical structure factor for two representative spin values $S = 0.12, 0.15$ at which a quantum spin liquid (QSL) can be achieved, and for various values of \Cuthree coupling $J_{10}$ for a system size of $l=12$ with $4032$ spins, showing the proximity of the phase transition between a quantum spin liquid state and its Bose condensate counterpart. 

\begin{figure}
  \includegraphics[width = 1.0\columnwidth]{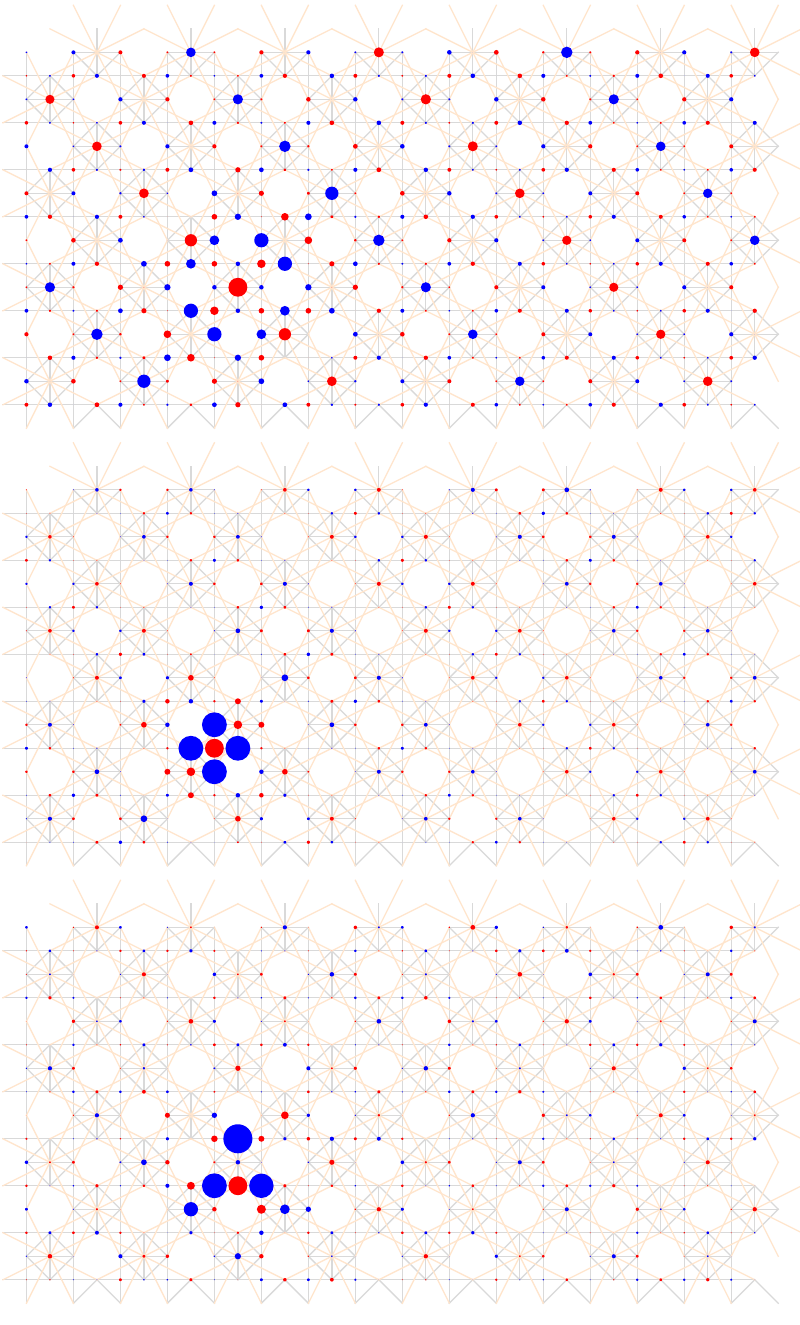}
  \caption{Real space (equal-time) spin-spin correlations considering 3 reference sites: (top) \Cuthree spin, (middle) corner of triangles and (bottom) corner of square plaquette. The parameters are the ones extracted in the main text, $S\simeq 0.366$ here, for which the sum rule of the static structure factor is satisfied~\cite{Lugan-2022}. The system size is $n_u \times 2 \times 6 \times 6$. The colors correspond to positive (blue) and negative (red) correlations. The reference site is always the largest red disk of the figure.}
  \label{fig:bzsbmft}
\end{figure}

As seen in the fermion approaches, the effect of projecting the wave function onto exact physical states increases the quantum fluctuations and helps the system to remain disordered even in the presence of the \Cuthree atoms. In the Schwinger boson mean field theory, since magnetic orders are more competitive by construction, they are favoured at $S=1/2$. Thus, in order to reach the QSL, one has to reduce the spin value. In Fig. 4 of the main text, we can see that the Bose condensation arises on the \Cuthree spins while the others on the square-kagome lattice remains mainly disordered. This is reflected by a clear gap in the lower panels between the condensed branches and the excitations in the continuum. To have a better view of this feature, we have plotted in Fig.~\ref{fig:bzsbmft} the real-space (equal-time) spin-spin correlations using three different reference sites on a \Cuthree spin (top panel), on a corner of a triangle (middle panel) and on a corner of a square plaquette (bottom panel).  

As one can see, a clear magnetic order appears on the \Cuthree spins while the ones on the remaining square-kagome lattice are disordered. This last feature is in good agreement with the observations of increased correlations between the \Cuthree sites in PMFRG, although no order could be detected in the accessible temperature regime.

\bibliography{squarekagome}